\documentstyle[multicol,aps,epsfig,amstex,amssymb]{revtex}
\begin{document}
\draft
\title{Dynamics of polymeric manifolds in melts: the Hartree approximation}
\author{V.G.Rostiashvili$^{(a,b)}$, M. Rehkopf$^{(a)}$ and T.A.
  Vilgis$^{(a)}$} 
\address{$^{(a)}$Max-Planck-Institut f\"ur Polymerforschung,
  Postfach 3148, D-55021 Mainz, Germany} 
\address{$^{(b)}$
  Chemical Physics, Russian Academy of Science, 142432, Chernogolovka, Moscow
  region, Russia}
\maketitle
\begin{abstract}
The Martin-Siggia-Rose functional technique and
the selfconsistent Hartree approximation is applied to the
dynamics of a $D$-dimensional manifold in a melt of similar manifolds. The generalized Rouse
equation is derived and its static and dynamic properties are studied. The
static upper critical dimension, $d_{\rm uc}=2D/(2-D)$, discriminates between
Gaussian (or screened) and non-Gaussian regimes, whereas its dynamical
counterpart, ${\tilde d}_{\rm uc}=
2d_{\rm uc}$, discriminates between Rouse- and
renormalized-Rouse behavior. The Rouse modes correlation function in a
stretched exponential form and the dynamical
exponents are calculated explicitly. The special case of linear chains $D=1$
shows agreement with Monte-Carlo-Simulations.
\end{abstract}

\begin{multicols}{2}
Much attention has been paid recently to the theory of the dynamical
behavior of polymers (or generally speaking polymeric manifolds) and
flux-lines in a quenched disordered random
medium \cite{1} or in a melt
\cite{2,3,4}. Technically either the
projection formalism and the mode-coupling approximation \cite{2,3} or the
selfconsistent Hartree approximation (HA)\cite{1,4} have been used for the derivation of the
equations of motion. 

In this paper, the HA is used to investigate the static and
mainly the dynamic properties of a
polymeric $D$ - dimensional manifold (or a fractal) in the
melt of the same manifolds. The generalized Rouse equation (GRE), which we
derived, reproduce in the static limit the screening and saturation
of $D$ - dimensional manifolds
\cite{7} in a different way. The whole dynamical consideration results in a
subdiffusive behavior and exponents, which are confirmed for the 3-dimensional
melt by MC and MD simulations \cite{8,8',8''}. We should stress,
that the manifolds in our consideration are crossable, so
that entanglements cannot occur and reptation dynamics is not considered.
The dynamics considered here corresponds to chains below the critical molecular weight:
entanglements are not of importance for the dynamic behavior \cite{9}.
We describe the manifolds below only in terms of
connectivity and excluded volume. The connectivity defines the $D$
- dimensional subspace
which is embedded in the Euclidean space of $d$
dimension. The model we have chosen allows
the interpolation between linear polymer chains, which correspond to $D=1$, and
tethered membranes ($D=2$). By analytic continuation to rational numbers of the
spectral dimension statements on polymeric fractals can be made. In a series of papers \cite{7}, we
have considered the different regimes in static scaling. In the present letter
we place emphasis on the dynamics. We will show below, that a new
dynamical regime for the motion
of the manifold segments appears.

Let us start with  a melt of $D$-dimensional manifolds in a $d$-dimensional
space. The test manifold is represented by the $d$-dimensional
vector ${\bf R}({\vec x},t)$ with the $D$-dimensional vector ${\vec x}$
of the internal coordinates. In
the same way the manifolds of the surrounding matrix are specified by ${\bf
  r}^{(p)}({\vec x},t)$ ($p=1,2,\ldots,M$). We have chosen the notation in
such a way,
that the boldfaced characters describe the external degrees of freedom in
Euclidian $d$ - dimensional space, whereas the arrow hatted vectors correspond
to the internal $D$ - dimensional space. The model of the melt of $M$
(monodisperse) tethered manifolds
used in the following is based on the generalized Edwards Hamiltonian,
\begin{multline}
{\cal H} = \frac{Td}{2l^{2}}
\sum_{p=1}^{M}
\int d^{D}x
\left( \nabla_{\vec x}{\bf r}^{(p)}{({\vec x})}\right)^{2}+\\
+\frac 12
\sum_{p,p'=1}^{M}
\int d^{D}x\int d^{D}x'
V\left({\bf r}^{(p)}(\vec x) -{\bf r}^{(p')}(\vec x')\right)
\end{multline}
In this melt an additional (test) manifold is immersed which is described by
the variables ${\bf R}({\vec x})$. The number of  monomers along one side of
the manifold is $N$ and limits the $\vec x$ - integration.

The corresponding Langevin equations in Cartesian components $j$ for the
test chain  has the form
\begin{multline}
\xi_{0}\frac{\partial}{\partial t}R_{j}({\vec
  x},t)-\epsilon\Delta_{x}R_{j}({\vec x},t)\\
+\frac{\delta}{\delta R_{j}({\vec
    x},t)}\int d^{D}x'\:V\left[{\bf R}({\vec x},t)-{\bf R}({\vec x'},t)
\right]\\
+\frac{\delta}{\delta R_{j}({\vec
    x},t)}\sum_{p=1}^{M}\int d^{D}x'\:V\left[{\bf R}({\vec x},t)-{\bf
    r}^{(p)}({\vec x'},t)\right]\\
=f_{j}({\vec x},t)
\end{multline}
and  similarly for all other polymers in the melt
\begin{multline}
\xi_{0}\frac{\partial}{\partial t}r^{(p)}_{j}({\vec
  x},t)-\epsilon\Delta_{x}r^{(p)}_{j}({\vec x},t)\\
+\frac{\delta}{\delta r^{(p)}_{j}({\vec
    x},t)}\int d^{D}x'\:V\left[{\bf r}^{(p)}({\vec x},t)-{\bf R}({\vec
x'},t)\right]\\
+\frac{\delta}{\delta r^{(p)}_{j}({\vec
    x},t)}\sum_{m=1}^{M}\int d^{D}x'\:V\left[{\bf r}^{(p)}({\vec x},t)-{\bf
    r}^{(m)}({\vec x'},t)\right]\\
={\tilde f}_{j}({\vec x},t)
\end{multline}
where $\xi_{0}$ is the bare friction coefficient, $\epsilon=Td/l^{2}$ the
elastic modulus with the Kuhn segment length $l$, $V(\cdots)$ the
excluded volume interaction function, $\Delta_{x}$ denotes a
$D$-dimensional Laplacian and the random forces $f_{j}$ and ${\tilde f}_{j}$ 
have the
standard Gaussian distribution.

We find it more convenient to reformulate the Langevin problem (2,3) in the  
MSR-functional integral representation \cite{4}. This representation
is especially useful for performing  transformations to
collective variables or
integration over a subset of variables. In our case we introduce the
matrix density $\rho({\bf r},t)$ and the response field density $\pi({\bf
  r},t)$
\begin{eqnarray}
\rho({\bf r},t)&=&\sum_{p=1}^{M}\int d^{D}x\:\delta({\bf r}-{\bf r}^{(p)}({\vec
  x},t))\\
\pi({\bf r},t)&=&\sum_{p=1}^{M}\sum_{j=1}^{d}\int d^{D}x\:i{\hat
  r}_{j}^{(p)}({\vec x},t)\nabla_{j}\delta({\bf r}-{\bf r}^{(p)}({\vec
  x},t))
\end{eqnarray}
In ref.~\cite{12} the first systematic expansion of the effective action in the
MSR-functional integral in terms of $\rho$ and $\pi$ was given.

The aim now is to integrate over the matrix variables (4,5). To do this, we
make the expansion of the effective Action up to the 2-nd order with respect
to $\rho$ and $\pi$, which corresponds to the random phase approximation
(RPA). After performing the (Gaussian) functional integration all information
about the matrix is comprised into the RPA correlation $S_{00}({\bf k},t)$ and
response $S_{01}({\bf k},t)$ functions \cite{12}.

The resulting Action includes still the test manifold variables in a highly
non-linear way. In order to handle it we use the  Hartree-type approximation and also take into account the
fluctuation-dissipation theorem for both, the test manifold and the matrix
variables. This strategy leads (see for details in \cite{4}) to the following GRE
\begin{multline}
\xi_{0}\frac{\partial}{\partial t}R_{j}({\vec
  x},t)\\
+\int d^{D}x'\int_{0}^{t}dt'\Gamma({\vec x},{\vec
  x}';t-t')\frac{\partial}{\partial t'}R_{j}({\vec
  x}',t')\\
-\int d^{D}x'\Omega({\vec x},{\vec x}')R_{j}({\vec
  x}',t)={\cal F}_{j}({\vec x},t)\label{GRE}
\end{multline}
with the memory function
\begin{multline}
\Gamma({\vec x},{\vec  x}';t)=\frac{1}{T}\int
\frac{d^{d}k}{(2\pi)^{d}}k^{2}|V({\bf k})|^{2}\\
\times F({\bf k};{\vec x},{\vec
  x}';t)S_{00}({\bf k},t)\label{Gamma}
\end{multline}
(where $F({\bf k};{\vec x},{\vec x}';t)$ is the test manifold density-density
correlator), the effective static elastic susceptibility
\begin{multline}
\Omega({\vec x},{\vec x}')=\epsilon\delta({\vec x}-{\vec x}')\Delta_{x}-\int
\frac{d^{d}k}{(2\pi)^{d}}k^{2}|{\cal V}({\bf k})|^{2}\\
\times{\Big [}F_{\rm st}({\bf k};{\vec
    x},{\vec x}')-\delta({\vec x}-{\vec x}')\int d^{D}x''F_{\rm st}({\bf k};{\vec
    x},{\vec x}'')
    ){\Big ]}\label{scrV}
\end{multline}
and the random force has the correlator
\begin{multline}
\left<{\cal F}_{i}({\vec x},t){\cal F}_{j}({\vec
    x}',t')\right>=2T\delta_{ij}{\Big [}\xi_{0}\delta({\vec x}-{\vec
    x}')\delta(t-t')\\
+\theta(t-t')\Gamma({\vec x},{\vec x}';t-t'){\Big ]}
\end{multline}
In eq.~(\ref{scrV}) the effective potential
\begin{equation}
{\cal V}({\bf k})=V({\bf k})\left[1-V({\bf k})S_{\rm st}({\bf k})/T\right]
\end{equation}
gains the standard screened form \cite{9}
\begin{equation}
{\cal V}({\bf k})=V({\bf k})\left[1+V({\bf
    k})F_{\rm st}^{(0)}({\bf k})/T\right]^{-1}\label{scrV2}
\end{equation}
(where $F_{\rm st}^{(0)}({\bf k})$ is the free system correlator) if the standard  
RPA-result is
used for the static correlator $S_{\rm st}({\bf k})$.

It is an important point that we treat on an equal footing both, the static and
dynamic parts of the GRE (\ref{GRE}). Let us start from the static behavior
of eqs.~(\ref{GRE})-(\ref{scrV2}). The static limit of these equations for
the Rouse mode correlator, $C({\vec p})=\left<{\bf R}({\vec
    p})\cdot{\bf R}(-{\vec p})\right>$, yields the Dyson-like form
\begin{equation}
C({\vec p})=\frac{d}{{\cal N}\left[\frac{d}{l^{2}}\left(\frac{2\pi{\vec
          p}}{N}\right)^{2}+\Sigma({\vec p})\right]}\label{Dyson}
\end{equation}
where ${\cal N}=N^{D}$ is the total number of monomers in the manifold and the
"self-energy" is given by
\begin{multline}
\Sigma({\vec
p})=\int\frac{d^{d}k}{(2\pi)^{d}}k^{2}\frac{V({\bf k})/T}{1+V({\bf k})F_{\rm
st}^{(0)}({\bf k})/T}\\
\times{\cal N}\left[F_{\rm st}({\bf k};{\vec
    p})-F_{\rm st}({\bf k};{\vec p}={\vec 0})\right]\label{self}
\end{multline}
The test manifold static correlator in eq.~(\ref{Dyson}) is parameterized by the
wandering exponent $\zeta$ in such a way
\begin{equation}
F_{\rm st}({\bf k};{\vec p})=\frac{1}{{\cal N}}\int
d^{D}x \exp \left(-\frac{k^{2}l^{2}}{2d}x^{2\zeta}-i
\frac{2\pi}{N}{\vec x}\cdot{\vec
    p}\right) \label{zeta}
\end{equation}
In its turn the exponent $\zeta$ is determined by the correlator $C({\vec p})$:
\begin{equation}
Q_{\rm st}({\vec x})=\int d^{D}p\left[1-e^{-i\frac{2\pi}{N}{\vec p}\cdot{\vec
      x}}\right]C({\vec p})\propto x^{2\zeta}\label{Q}
\end{equation}
The system of eqs.~(\ref{Dyson})-(\ref{Q}) should be analyzed
self-consistently. Straightforward calculations yield
\begin{equation}
\Sigma({\vec p})=-c_{1}\left(\frac{\pi{|\vec
      p|}}{N}\right)^{\zeta(d+2)}-c_{2}\left(\frac{\pi{|\vec
      p|}}{N}\right)^{\zeta(d+d_{\rm f}^{0}+2)-D}\label{Sigma2}
\end{equation}
with the Gaussian fractal dimension $d_{\rm f}^{0}=2D/(2-D)$ \cite{7,13} and
$c_{1}$, $c_{2}$ are some constants depending on $d$ and $D$. Physically, the
condition for the exponent $\zeta$ comes from the balance between the
entropic and the interaction terms in the denominator of eq.~(\ref{Dyson}). The
analysis shows that the only way to satisfy the eqs.~(\ref{Dyson})-(\ref{Q})
is to impose the exponents in eq.~(\ref{Sigma2}) the condition:
$\zeta(d+2)=\zeta(d+d_{\rm f}^{0}+2)-D\geq 2$. This holds at $d\geq
d_{\rm uc}=2D/(2-D)$ and the only solution is the Gaussian one with
$\zeta=\zeta_{0}=(2-D)/2$. Besides that, the necessary condition $d_{\rm f}^{0}<d$
immediately implies $D<D_{s}=2d/(2+d)$. The {\it upper critical
dimension} $d_{\rm uc}$ in a melt and the {\it spectral critical dimension}
$D_{s}$ was discussed first in
\cite{7}. At $d<d_{\rm uc}$ the interaction term in eq.~(\ref{Dyson})
overwhelms, $|\Sigma({\vec p})|\gg (2\pi{\vec p}/N)^{2}$, and the system become
unstable. The manifold is saturated in a melt, i.e. it loses its fractal nature
and becomes compact \cite{7}.

We now consider the dynamics at $d\geq d_{\rm uc}$. There are two dynamic
exponents, $z$ and $w$. The exponent $z$ measures the time dependence of a
monomer displacement, i.e.
\begin{equation}
Q(t)=\int d^{D}p\int_{a-i\infty}^{a+i\infty}\frac{ds}{2\pi i}
\left[1-e^{st}\right]C({\vec p},s)\propto
t^{2z}\label{z}
\end{equation}

and the exponent $w$ measures the same for the center of mass
\begin{equation}
Q_{\rm c.m.}(t)=\lim_{{\vec p}\rightarrow
0}\int_{a-i\infty}^{a+i\infty}\frac{ds}{2\pi i}\left[1-e^{st}\right]C({\vec  
p},s)\propto
t^{w}\label{w}
\end{equation}
In eqs.~(\ref{z})-(\ref{w}) $C({\vec p},s)=\left<|{\bf R}({\vec
    p},s)|^{2}\right>$ is the Rouse-Laplace component of the correlator
$C({\vec x},t)$. The formal solution of eqs.~(\ref{GRE})-(\ref{scrV}) for $C({\vec p},s)$
is given by
\begin{equation}
C({\vec p},s)=\frac{C_{\rm st}({\vec p})}{s+\frac{\epsilon\left(\frac{2\pi  
{\vec p}}{N}\right)^{2}}{\xi_{0}+{\cal N}\Gamma({\vec p},s)}}\label{fsol}
\end{equation}
where $C_{\rm st}({\vec p})=l^{2}(N/2\pi{\vec p})^{2}/{\cal N}$ and
$\Gamma({\vec p},s)$ is the Rouse-Laplace transformation of the memory function
(\ref{Gamma}).

In the RPA the matrix density correlator $S_{00}({\bf k},t)$ is well
approximated by
\begin{multline}
S_{00}({\bf k},t)=S_{\rm st}({\bf k})\\
\times\left\{\begin{array}{r@{\quad,\quad}l}
\exp\left\{-k^{2}D_{\rm coop}({\bf k})t\right\}\:\: & (kl)^{d_{\rm f}^{0}}{\cal  
N}\ll 1\\
\exp\left\{-\frac{k^{2}l^{2}}{2d}\left(\frac{t}{\tau_{0}}\right)^{2z_{0}}\right\}  
& (kl)^{d_{\rm f}^{0}}{\cal N}\gg 1\end{array}\right.\label{S00}
\end{multline}
where the cooperative diffusion coefficient $D_{\rm coop}({\bf k})=\rho V({\bf
  k})/\xi_{0}$, $\tau_{0}=\xi_{0}l^{2}/Td$ and $z_{0}=(2-D)/4$ is the Gaussian
$z$-exponent. The corresponding Ansatz for the test manifold yields
\begin{multline}
F({\bf k};{\vec x};t)=F_{\rm st}({\bf k},{\vec
  x})\\
\times\left\{\begin{array}{r@{\quad,\quad}l}\exp\left\{-k^{2}D_{G}t\right\}\quad\quad\:\:  
& (kl)^{d_{\rm f}^{0}}{\cal N}\ll 1\\
\exp\left\{-\frac{k^{2}l^{2}}{2d}\left(\frac{t}{\tau_{0}}\right)^{2z}\right\}
& (kl)^{d_{\rm f}^{0}}{\cal N}\gg 1\end{array}\right.\label{Ff}
\end{multline}
where $D_{G}$ is the self-diffusion coefficient.

(i) By making use of
eqs.~(\ref{S00})-(\ref{Ff}) and eq.~(\ref{Gamma}) in the limit
$(kl)^{d_{f}^{o}}{\cal N}\gg 1$ one can derive the result
\begin{equation}
{\cal N}\Gamma({\bf
  p},s)={\text const.}\left(\frac{1}{\tau_{0}s}\right)^{1-\beta}\label{Gammab}
\end{equation}
with
\begin{equation}
\beta=z_{0}(d-d_{\rm uc}+2)
\end{equation}
In the derivation of eq.~(\ref{Gammab}) we also used the assumption $l^{2}\ll
Q(t)\ll l^{2}{\cal N}^{2/d_{\rm f}^{0}}$. The condition $\beta<1$ immediately
defines the {\it dynamical upper critical dimension}
\begin{equation}
{\tilde d}_{\rm uc}=\frac{4D}{2-D}\label{dd}=2d_{\rm uc}
\end{equation}
i.e. the dimension above which the manifold has the simple Rouse behavior, at
$d={\tilde d}_{\rm uc}$ one can call it the marginal Rouse behavior and only at
$d<{\tilde d}_{\rm uc}$ the dynamic exponents $z$ and $w$ are renormalized.

At $d<{\tilde d}_{\rm uc}$ eqs.~(\ref{fsol}), after inverse Laplace
transformation, yields
\begin{multline}
C({\vec p},t)=C_{\rm st}({\vec
  p})\\
\times\sum_{m=0}^{\infty}\left[-\epsilon A\left(\frac{2\pi {\vec
p}}{N}\right)^{2}\left(\frac{t}{\tau_{0}}\right)^{\beta}\right]^{m}{\Big
/}\Gamma(m\beta+1)\label{strex}
\end{multline}
where $A=\left[\beta|V(k=0)|^{2}S_{\rm st}(k=0)/l^{d+2}\right]^{-1}$ and
$\Gamma(x)$ is the gamma-function. The eq.~(\ref{strex}) is very close to the
stretched exponential form found by the MC- and MD-simulations \cite{8}. The
eq.~(\ref{strex}) was actually calculated in the limit $p\rightarrow 0$, and we
can use it first of all to comparison with simulation results on the center of
mass mean square displacement. By using eq.~(\ref{strex}) in eq.~(18) we
obtain\begin{eqnarray}
Q_{{\rm cm}}(t)=\frac{{\cal D}_{0}}{{\cal
    N}}\left(\frac{t}{\tau_{0}}\right)^{w}\label{26}
\end{eqnarray}
where ${\cal D}_{0}=l^{2}\epsilon A/\Gamma(\beta+1)$ and
\begin{eqnarray}
w=\beta=z_{0}(d-d_{{\rm uc}}+2)\label{27}
\end{eqnarray}
By the same way from eq.~(\ref{z}) we can derive:
\begin{eqnarray}
z=z_{0}\beta=z_{0}^{2}(d-d_{{\rm uc}}+2)\label{28}
\end{eqnarray}
In ref.~[3] this renormalized dynamics was formally used as a projected
dynamics for the polymer mode coupling approximation (PMCA) and eventually
leads to the GRE which, as the authors claim, can describe the entangled
dynamics. We will argue in an extended paper \cite{10'} that this is a result
of misinterpretation of the GRE.

(ii) If we assume that the main contribution to the integral (7) comes from
  the small wave vectors, $(kl)^{d_{f}^{0}}{\cal N}\ll 1$, then we arrive at 
\begin{eqnarray}
{\cal N}\Gamma({\vec p},s)\propto s^{(d-d_{{\rm uc}})/2}\label{29}
\end{eqnarray}
Since $d>d_{{\rm uc}}$, the simple Rouse behavior in the small wave vector
regime does not change.

At the large
displacement regime, $R_{\rm G}^{2}\ll Q_{\rm c.m}(t)$, one should expect a
simple diffusive behavior:
\begin{equation}
Q_{\rm c.m.}(t)=dD_{\rm G}t\label{diff1}
\end{equation}
with
\begin{equation}
D_{\rm
  G}=T/{\cal N}[\xi_{0}+{\cal N}\Gamma(p=0;s=0)]\label{diff2}
\end{equation}
Now the problem is how
$D_{\rm G}$ depends from ${\cal N}$? One can assume that for this case only
the small wave vectors, $(kl)^{d_{\rm f}^{0}}{\cal N}\ll 1$, are relevant,
i.e. the dynamics of the matrix is driven by the cooperative diffusion
coefficient $D_{\rm coop}$ and the dynamics of the test manifold by the self-
diffusion one $D_{\rm G}$ (see eq.~(20-21)). Since in any way $D_{\rm
  coop}\gg D_{\rm G}$, the calculation yields
\begin{equation}
{\cal N}\Gamma(p=0;s=0)\propto[D_{\rm coop}]^{-1}{\cal  
N}^{(1-\frac{d}{d_{\rm f}^{0}})}
\end{equation}
\begin{figure}
\epsfig{file=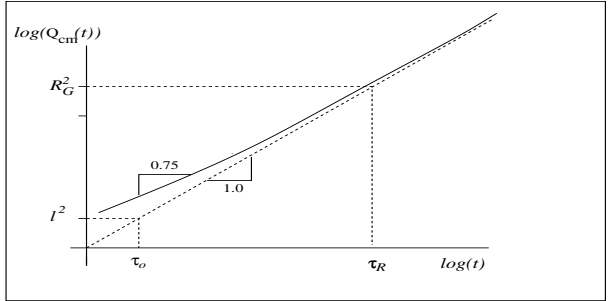,width=8cm,height=4cm}
\caption{A schematic plot of $Q_{\rm cm}(t)$ for the simple Rouse 
(dashed line) and the renormalized Rouse (solid line) dynamics}
\end{figure}
But $D_{\rm coop}={\cal O}(N^{0})$ and $d_{\rm f}^{0}<d$, then
$\Gamma(p=0;s=0)\rightarrow 0$ at ${\cal N}\rightarrow\infty$. As a result
$D_{\rm G}=T/{\cal N}\xi_{0}$, i.e. the simple Rouse result does not change. 

In Fig.~1 we have summarized the overall schematic behavior for $Q_{{\rm
    cm}}(t)$. At the relatively short times, $\tau_{0}<t\ll \tau_{R}$, and
displacements, $l^{2}<Q_{{\rm cm}}(t)\ll R_{G}^{2}$, the test chain dynamics
is mainly determined by the fluctuations from the interval
$(kl)^{d_{f}^{o}}{\cal N}\gg1$. As a result the renormalized Rouseian behavior
dominates and e.g. for the melt of polymer chains ($D=1$) in the 3 -
dim. space the exponent $w=3/4=0.75$. In the opposite limit, $\tau_{R}\ll t$
and $R_{G}^{2}\ll Q_{{\rm cm}}(t)$, the long wavelength fluctuations
$(kl)^{d_{f}^{o}}{\cal N}\ll 1$ are relevant and the melt almost does not
influence the test chain: The simple Rouse regime is recovered.

MC-simulations of the bond-fluctuation model \cite{8} as well as the
MD-simulations \cite{8'} of the athermal melt have been undertaken. Recently also the static and
dynamic properties of a realistic polyethylene melt have been studied
\cite{8''}. Both in MC and MD simulations a slowed down motion at intermediate
times for the center of mass is clearly observable. It was found e.g. that for
the chain length $N=200$ at the relatively short time $Q_{{\rm cm}}(t)\propto
t^{w}$ with $w=0.8$ (instead of $w=1$) in \cite{8} and $w=0.71$ in
\cite{8'}. This deviation from the simple Rouse regime also occurs for short
chains ($N<N_{e}$) which clearly are not entangled \cite{8''}.

The best test of the renormalized Rouse dynamics predictions would be the
simulation of rather long crossable (to avoid reptation) chains but still with
an excluded volume interaction. In a recent MC-simulation \cite{14} the
statics and dynamics of such melts have been studied. Unfortunately in
\cite{14} the plot $Q_{\rm cm}(t)$ is not given explicitly, i.e. it stays
unclear from this simulation how the mode $p\rightarrow 0$ is renormalized.

In summary, we have shown that by using MSR-functional technique and the
Hartree approximation the GRE for a $D$-dimensional manifold in the melt of
similar manifolds
can be derived. In this equation the static and dynamic parts are treated on
an equal footing. Besides the static upper critical dimension,
$d_{\rm uc}=2D/(2-D)$, its dynamical counterpart,
${\tilde d}_{\rm uc}=2d_{\rm uc}$
was found (such that at $d_{\rm uc}<d<{\tilde d}_{\rm uc}$ the manifold is
Gaussian but renormalized-Rouseian). We have calculated the dynamical
exponents, $w$ and
$z$, and have explained some novel computer simulation findings.

We have benefited from discussions with J. Baschnagel, K. Binder, K. Kremer,
W. Paul and  
T.B. Liverpool. We
are greatful  to  the Sonderforschungsbereich 262 and
the Bundesminister f\"ur Bildung und Forschung for finacial support.

\end{multicols}
\end{document}